%% file: main.tex
\documentclass[11pt]{article}
\pdfoutput=1

\usepackage[T1]{fontenc}              
\usepackage[latin9]{inputenc}         

\usepackage[a4paper]{geometry}        

\usepackage{url}

\usepackage{booktabs}

\usepackage{graphicx} 
\usepackage{amssymb,amsfonts,amsmath}

\usepackage{natbib}
\usepackage{mathpazo}
\usepackage{microtype}

\input{abbreviations}

\begin{document}

\date{\today}

\title{
Effect Size Estimation  and  Misclassification Rate Based Variable Selection in 
Linear Discriminant Analysis 
}

\author{Bernd Klaus 
      \thanks{Institute for Medical Informatics,
      Statistics and Epidemiology,
      University of Leipzig,
      H\"artelstr. 16--18,
      D-04107 Leipzig, Germany,
      E-mail: bernd.klaus@uni-leipzig.de,
      Tel.: +49 341 97 16104,
      Fax: +49 341 97 16109 } 
}

\maketitle
\input{abstract}

\newpage

\input{body}

\input{acknowledgments}

\newpage

\bibliographystyle{apalike}
\bibliography{preamble,econ,genome,stats,array,sysbio,misc,molevol,med,entropy,Bernd}

\end{document}

%% file: abbreviations.tex
\newcommand\bdelta{\boldsymbol \delta}

\newcommand\bmu{\boldsymbol \mu}
\newcommand\bomega{\boldsymbol \omega}

\newcommand\bP{\boldsymbol P}

\newcommand\bS{\boldsymbol S}
\newcommand\bSigma{\boldsymbol \Sigma}

\newcommand\bV{\boldsymbol V}

\newcommand\bx{\boldsymbol x}

\newcommand\prob{\text{P}}

\newcommand\fdr{\text{fdr}}

\providecommand{\abs}[1]{\left\lvert#1\right\rvert}

\newcommand{\figcite}[1]{Fig.~\textbf{\ref{#1}}}
\newcommand{\eqcite}[1]{Eq.~\textbf{\ref{#1}}}
\newcommand{\tabcite}[1]{Tab.~\textbf{\ref{#1}}}

%% file: abstract.tex
\begin{abstract}

Supervised classifying of biological samples based on genetic information,
(e.g. gene expression profiles) is an important problem in biostatistics. In order
to find both accurate and interpretable classification rules variable selection is 
indispensable.

This article explores how an assessment of the  individual importance of variables 
(effect size estimation) can be used to perform variable selection.
I review recent effect size estimation approaches in the context of linear 
discriminant analysis (LDA) and propose a new conceptually simple 
effect size estimation method which is  at the same time computationally efficient.   

I then show how to use effect sizes to perform variable selection based on 
the misclassification rate which is the data independent expectation of the 
prediction error. Simulation studies and real data analyses illustrate that 
the proposed effect size estimation and variable selection methods are competitive. 
Particularly, they lead to both  compact and interpretable feature sets. \\

\textbf{Key Words:} correlation-adjusted $t$-score; effect size estimation; linear discriminant 
analysis; misclassification rate; variable selection 

\end{abstract}

%% file: body.tex
\section{Introduction}

Modern medical research has been revolutionized by the possibility of
characterizing diseases at a molecular level using microarrays. 
Classification of biological samples based on their gene expression continues
to be a field of active research. 
See e.g. \cite{Pang+2009, Cao+2011, WS2011} 
and \cite{Shao+2011}. Current reviews of the subject can be found in  \cite{SIR08, 
SDB08}  as well as in \cite{KSim2011}. 

In order to develop classifiers which are potentially useful for  molecular 
diagnostics it is important to construct them based on a selection of genes  
(variables)  strongly associated with the respective class labels (e.g. cancer and 
healthy tissue). These genes possess  a large effect size which is
generally measured by standardized differences. 

Three distinct but closely related  objectives need to be achieved
to identify a group of genes with high effect sizes \citep{AS2010, Matsui2011}:  
\begin{itemize}
\item[(i)] to establish a reliable variable ranking,
\item [(ii)] to provide a reasonable estimate of the effect size for each gene, and
\item[(iii)] to find a suitable  cutoff point that allows one to disregard 
(the usually large) number of noise-features.
\end{itemize}

Problems (ii) and (iii) are the main concerns of the current article. For the ranking 
problem (obj. (i)) I rely on correlation adjusted $t$--scores (a.k.a. ''cat`` -- scores) 
introduced by \cite{ZS09}. The cat--score  is a $t$--type statistic which takes 
correlation into  account and has been shown to induce a reliable variable ranking even in the
presence of correlation among the variables. I therefore use cat--scores  to
 obtain effect size estimates (obj. (ii)).  
Based on these estimates a nominal prediction error is computed.  It is dependent 
on the number of variables included.
Variable selection is then performed  (ob. (iii))  by determining 
the number of variables necessary  to achieve a certain nominal error level.
I choose to use linear discriminant analysis (LDA) as a classification method
 -- a simple yet very effective approach to  linear classification \citep{Hand06}. 

The approach of this paper is similar to that of \cite{Efr09} and \cite{DS07}.
However in contrast to  \cite{Efr09} my method applies to any number of classes
and allows empirical null modeling. In contrast to \cite{DS07} it does not need a computationally 
expensive greedy algorithm to select variables  due to the variable ranking 
performed beforehand. 

The article is organized as follows: I present basic theory on LDA 
in chapter \ref{LDA}, then I  obtain effect size estimates  based on cat--scores  
and compare them to other effect size estimation approaches 
in chapter \ref{effect}. Notably the \cite{Efr09} and \cite{Matsui2011} methods 
are presented in a unifying way using cat-- scores which sheds new light on their
similarities.
 Chapter \ref{VAR} shows how to perform 
variable ranking and selection combining methodology introduced in chapters 
\ref{LDA} and \ref{effect}. Results of the derived variable selection method on 
simulated and real data are then presented in chapter \ref{RES}. A discussion
concludes the article.



\section{Linear Discriminant Analysis (LDA) and its Misclassification Rate}\label{LDA}
\subsection{Linear discriminant analysis (LDA) and effect sizes}

LDA forms the basis of most classification algorithms currently employed, 
e.g. Nearest Shrunken Centroids commonly abbreviated as NSC, and 
also known as PAM, see \cite{THNC03}, Shrinkage 
Discriminant Analysis -- SDA, \cite{AS2010} -- and many more. 
It starts by assuming a mixture model 
for the $d$-dimensional data $\bx$
$$
f(\bx) = \sum_{k=1}^K \pi_k f(\bx | k) ,
$$
where each class $k$ is represented by a  multivariate normal 
density 
\begin{equation*}
f( \bx | k )  = (2 \pi)^{-d/2} | \bSigma|^{-1/2} \times 
 \exp\{  -\frac{1}{2} (\bx-\bmu_k)^T \bSigma^{-1} (\bx-\bmu_k) \} \, ,
\end{equation*}
with group--specific centroids $\bmu_k$ and a common covariance matrix $\bSigma$.
A sample $\bx$ is assigned to the class yielding the highest  LDA discriminant 
score defined as the log posterior probability $d_k^\text{LDA}(\bx) = \log\{\prob(k| \bx)\}$.
This score can be written as
\begin{equation}
d_k^\text{LDA}(\bx) = 
 \bmu_k^T \bSigma^{-1} \bx -\frac{1}{2} 
 \bmu_k^T \bSigma^{-1} \bmu_k  + \log(\pi_k) \,.
\label{eq:discrimLDA}
\end{equation}

The standard form of the LDA predictor function 
shown in \eqcite{eq:discrimLDA}
can be transformed into a scalar product which is given by 
\begin{gather}
\Delta_k^\text{LDA}(\bx) = 
\left(\bomega^{(k, \text{pool})}\right)^T \bdelta_k(\bx)
+ \log(\pi_k). \label{eq:multiclasslda}
\intertext{See \cite{AS2010} for details. In \eqcite{eq:multiclasslda} 
we have an inner product of Mahalanobis transformed variables (commonly called features )  $\bdelta(\bx)$ 
and a corresponding feature  weight vector $\bomega^{(k, \text{pool})}$ given by }
\bdelta_k(\bx) =  \bP^{-1/2} \bV^{-1/2}  \left( \bx - \frac{ \bmu_k+\bmu_\text{pool}}{2}\right) \, 
\label{eq:distfunc}
\intertext{and}
\bomega^{(k, \text{pool})} =  \bP^{-1/2} \bV^{-1/2} ( \bmu_k -  \bmu_\text{pool} )
\label{eq:featureweights}
\end{gather}
respectively.
In this equation  the pooled mean is calculated as
$\bmu_\text{pool} =  \sum_{k=1}^K \frac
{n_k}{n} \bmu_k$ and 
the covariance matrix $\bSigma$ is decomposed as: $\bSigma = \bV^{1/2} 
\bP \bV^{1/2}$, with a diagonal matrix containing the variances   $\bV = \text{diag}
\{\sigma^2_1, \ldots, \sigma^2_d\}$  and the correlation matrix $\bP = (\rho_{ij})$. 
Remarkably, both $\bomega^{(k, \text{pool})}$ and $\bdelta_k(\bx)$ are  vectors and not 
matrices. 

The decomposition in \eqcite{eq:multiclasslda} shows that 
$\bomega^{(k, \text{pool})}$  gives the influence of the \emph{transformed} 
variables $\bdelta(\bx)$   in prediction. \cite{ZS09} have shown that this
Mahalanobis--transformation leads to an improved ranking of the 
\emph{original} variables since it removes the effect of correlation. 
Thus, as in \cite{AS2010} the feature weights
$\bomega$ will serve a measure of variable importance and the terms variables 
and features will be used interchangeably in the following.

Additionally from \eqcite{eq:featureweights} it can be seen that the components of $
\bomega^{(k, \text{pool})}$ are  decorrelated and 
standardized differences (i.e. effect sizes) 
between the class $k$ and the ''pooled class`` \citep{Matsui2011}. This is readily generalized. 
The effect size vector $\bomega^{(k,l)}$ between any two classes $k$ and $l$ is 
defined as the  difference between the two respective feature weight vectors $\bomega^{(k, \text
{pool})}$ 
and $\bomega^{(l, \text{pool})}$
\begin{gather}
 \bomega^{(k,l)}  := \bomega^{(k, \text{pool})} - \bomega^{(l, \text{pool})}
 = \bP^{-1/2} \bV^{-1/2} ( \bmu_k-  \bmu_l )  \, . \label{eq:effect size}
\end{gather}
Note that $\bomega^{(k,l)}$ is up to the scale factor $(1/n_{k}+1/n_{l})^{-1/2}$ 
equivalent to the cat--score vector between the 
classes $k$ and $l$ on the population level, i.e. assuming known model parameters
\citep{ZS09}. Hence there is a close relationship between test statistics and effect 
sizes: The effect size is simply a sample size independent version of the test statistic.
The statistic is denoted by a ``cat'' subscript in this article, i.e. 
\begin{gather*}
\bomega^{(k,l)}_{\text{cat}} = (1/n_{k}+1/n_{l})^{-1/2}\bomega^{(k,l)}\, .
\end{gather*}

\subsection{The misclassification rate of linear discriminant analysis}

In this section I look at an unconditional (i.e. not depending on the data)
misclassification error of LDA on the population level. 
This quantity is called (unconditional) misclassification rate in the literature
\citep{DS07, Shao+2011}.

Let  $\bx^{(k)}$ be a sample  vector drawn from the multivariate normal distribution $N
(\bmu_k, 
\bSigma)$ associated with class $k$. In the LDA algorithm it is assigned to the  class 
yielding 
the highest score (\eqcite{eq:discrimLDA}). 
Using  the scalar product of \eqcite{eq:multiclasslda} a misclassification 
of $\bx^{(k)}$ occurs if 
 $[\bomega^{(k, \text{pool})}]^T\bdelta_{k}(\bx^{k}) +  \log(\pi_{k}) <  \max_{l}
 [\bomega^{(l, \text{pool})}]^T\bdelta_{l}(\bx^{(k)}) +  \log(\pi_{l})$. It is easily 
verified that this is equivalent to the condition

\begin{gather}
\min_{l \neq k} \frac{[\bomega^{(k,l)}]^{T} [\bP^{-1/2}\bV^{-1/2} \left(\bx^{(k)} - \frac{\bmu_{k} + 
\bmu_{l}}{2} \right)] 
+ \log \left( \frac{\pi_{k}}{\pi_{l}} \right)} 
{ \sqrt{[\bomega^{(k,l)}]^{T}[\bomega^{(k,l)}]}}  <0 
\notag \, .
 \intertext{Since $\bx^{(k)} \sim N(\bmu_k, \bSigma)$ holds for all $k \in \{1, \dotsc, K\}$, the  
expected probability of misclassifying a sample from class $k$ into a wrong
class $j \neq k $ can be deduced from the
above formula as:}
\prob(  j \neq k | k) = \Phi \biggl( - \min_{l \neq k} \frac{[\bomega^{(k,l)}]^T [\bomega^{(k,l)}] + 
2 \log\biggl(\frac{\pi_k}{\pi_l}\biggr) } {2 \sqrt{[\bomega^{(k,l)}]^{T}[\bomega^{(k,l)}]}}  
\biggr) \, . \notag  \label{eq:serror}
\intertext{This results in a misclassification rate (total error probability) of}
\prob(\text{error}) 
= \sum_{k=1}^{K} \prob( j \neq k | k)  \times \prob(  k)
= \sum_{k=1}^{K}  \Phi \biggl( - \min_{l \neq k} \frac{[\bomega^{(k,l)}]^T [\bomega^{(k,l)}] + 
2 \log\biggl(\frac{\pi_k}{\pi_l}\biggr) } {2 \sqrt{[\bomega^{(k,l)}]^{T}[\bomega^{(k,l)}]}}  
\biggr) \times \pi_{k} \, . \label{eq:terror}
\end{gather}

\section{Effect Size Estimation}
\label{effect}
For two given classes $k$ and $l$ a feature $i$ with a large corresponding 
effect size $\omega_i^{(k,l)}$ is most influential in differentiating between
class $k$ and $l$. However a ``naive'' estimation of $\omega_{i}^{(k,l)} $
(e.g. estimation by  plug-in estimates) suffers from the so called 
``selection bias'': estimates of $\omega_{i}^{(k,l)}$ are  biased upwards
in general. For example an estimated effect size of 1.5 based on $t$--scores 
might correspond to a true effect size of 0.7, see  \figcite{fig:simulation}.
Therefore  reliable estimates of $\omega_{i}^{(k,l)} $ are needed
in order to furnish a good estimate  of \eqcite{eq:terror}.

\subsection{Three empirical Bayes approaches} \label{sec:empBay}
Bayesian approaches are ``immune'' to selection effects \citep{Dawid1994, 
Senn2008}. Thus, both \cite{Efr09} and \cite{Matsui2011} 
employ empirical Bayes estimates to tackle the estimation of effect sizes. 

I am going to present their ideas in a unified way using cat--scores. This will
show similarities between the two methods that are not readily apparent from
studying the two original papers. Therefore a both methods are presented 
in considerable detail to clearly demonstrate the conceptual overlap
between them. This will also help to indicate their respective weaknesses.

Furthermore, the current section can be read as concise and yet comprehensive
review of both methods which can be of great help to the interested reader. 
The empirical Bayes estimator presented in section \ref{sec:simplEB} is an attempt
to combine the strengths of both approaches while adressing their shortcomings.

Let $k$ and $l$ be any two classes. For the sake of simplicity the feature index $i$  
($i \in \{1, \dotsc, d\}$) will be dropped in the upcoming subsections.

\subsubsection{\cite{Efr09}}

\cite{Efr09} begins with transforming the statistics $\omega^{(k,l)}_{ \text{cat}}$ 
 into $z$--scores via a $t$--distribution with $n_{l} + n_{k}-2$ 
degrees of freedom:
\begin{gather}
z = \Phi^{-1}\left(F_{n_{l} + n_{k}-2}(\omega^{(k,l)}_{ \text{cat}})\right) \, ,   
\notag
\intertext{
where $F_{n_{l} + n_{k}-2}$ denotes the distribution function of a
$t$--distribution with $n_{l} + n_{k}-2$ degrees of freedom. He then assumes
a prior density $g$ on $\omega^{(k,l)}_{ \text{cat}}$ given by the 
mixture}
g(\omega^{(k,l)}_{ \text{cat}}) = \eta_0 I_0(\omega^{(k,l)}_{ \text{cat}}) + (1-\eta_0)g_A
(\omega^{(k,l)}_{ \text{cat}}) \, , \label{eq:priorEf}
\intertext{where $I_0$ is a delta-function at 0 and $\eta_0$ the proportion
of genes having a true effect size of zero. The alternative group, i.e. the 
nonzero effect sizes are represented by $g_A$. In the following I will
 in general abbreviate conditioning on the alternative group
with an ''$A$`` subscript. The statistic $z$ is assumed to
be distributed as}
z|\omega^{(k,l)}_{ \text{cat}}  \sim N(\omega^{(k,l)}_{ \text{cat}},1).  \notag
\intertext{Together with \eqcite{eq:priorEf} this results in
 the following mixture model for $z$}
f(z)  =  \eta_0 \varphi(z) + (1-\eta_0) f_A(z) \, , \label{eq:EfMix}
\intertext{where $\varphi(z)$ is the normal distribution density and $f_A$ is a
mixture of the densities $\varphi(z -\omega^{(k,l)}_{ \text{cat}})$:}
f_A(z) = \int_{-\infty}^{\infty} \varphi(z -\omega^{(k,l)}_{ \text{cat}}) g_A(\omega^{(k,l)}_{ \text{cat}}) \, d \omega^{(k,l)}_{ \text{cat}}. \notag
\intertext{\eqcite{eq:EfMix} is a typical case of two-group mixture model common in multiple testing situations. It consists of a theoretical (i.e. no additional parameters) ``null'' model 
$f_0 = \varphi$ and an alternative component $f_A$ from which the 
``interesting'' cases are assumed to be drawn \citep{Efr08a}. In order to present
the ideas of both \cite{Matsui2011} and \cite{Efr09} in a unified fashion 
I start with computing the posterior density conditioned on the alternative, i.e. 
$f( \omega^{(k,l)}_{ \text{cat}}| z, z  \in \text{``alternative''} )
= f( \omega^{(k,l)}_{ \text{cat}}| z, \omega^{(k,l)}_{ \text{cat}} \neq 0 )
$.
As introduced above the ''$A$`` subscript indicates conditioning 
on the alternative,  so that $f_A( \omega^{(k,l)}_{ \text{cat}}| z)
 = f( \omega^{(k,l)}_{ \text{cat}}| z, z  \in \text{``alternative''} )$.  
Finally, using Bayes' rule  this density can be computed as}
f_A( \omega^{(k,l)}_{ \text{cat}}| z)
=  \frac{ f_A( z |  \omega^{(k,l)}_{ \text{cat}} ) \cdot g_A ( \omega^{(k,l)}_{ \text{cat}} ) }
{ f_A( z )} \notag \\
= \exp (\omega^{(k,l)}_{ \text{cat}} z - \log\{f_A(z) / \varphi(z)\} )
 [\exp\{-(\omega^{(k,l)}_{ \text{cat}})^2/2)\}]
g_A(\omega^{(k,l)}_{ \text{cat}}) \,. \notag 
\intertext{It has the form of a natural exponential family with natural
parameter  $\omega^{(k,l)}_{ \text{cat}}$, sufficient statistic $z$ and cumulant 
generating function $\log\{f_A(z) / \varphi(z)\} = \log\{[(1-\fdr(z)) / \fdr(z)]\} \cdot \eta_0(1-
\eta_0) \} $, where}
\fdr(z) =  \prob(\text{``null''} | z)  =  \eta_0 \frac{ \varphi(z)}{f(z)} 
= \eta_0 \frac{ f_0(z)}{f(z)} \, \label{eq:fdrDef}
\intertext{is the local false discovery rate \citep{Efr08a}. 
Conditional on the alternative this  leads to an effect size estimate 
of the simple form}
E_A\left(\omega^{(k,l)} | z\right)
 = -(1/n_{l}+1/n_{k})^{1/2}\frac{d}{d z} \log\left(\frac{1-\fdr(z)}
{\fdr(z)} \frac{\eta_0}{1-\eta_0} \right) \,. \label{eq:conEff} %
\intertext{Since by \eqcite{eq:fdrDef} the relationship $\prob(\text{``alternative''} | z)  = 1 -
\prob(\text{``null''} | z)
= 1 - \fdr(z)$ holds, the unconditional effect size estimate  is:  }
E\left(\omega^{(k,l)} | z\right)  \notag
= E_A\{\omega^{(k,l)} | z\} \{1 - \fdr(z)\} \\ 
= -(1/n_{l}+1/n_{k})^{1/2}\frac{d}{d z} \log\left\{\frac{1-\fdr(z)}{\fdr(z)} 
\frac{\eta_0}{1-\eta_0} \right\}\{1 - \fdr(z)\}\, , \label{eq:unconEff}
\intertext{which after some further calculations becomes}
E\left(\omega^{(k,l)} | z\right) = -(1/n_{l}+1/n_{k})^{1/2}\frac{d}{d z} \log\{\fdr(z)\}\, . \label{eq:EffEf}
\end{gather}
Note that if one used  an empirical null $N(0,\sigma^2)$ with estimated $\sigma$
as null density $f_0$ the connection to the natural exponential family would be lost. 
Then both the natural parameter  and the sufficient statistic would depend on $\sigma$. 

Unfortunately, in this case the elegant formula \eqref{eq:EffEf} no longer holds. This  basically is the
only downside of Efron's approach: It is conceptually simple and computationally 
efficient but it is not possible to include an additional variance parameter in the null 
model without ``destroying'' \eqcite{eq:EffEf}.

\subsubsection{\cite{Matsui2011}}

\cite{Matsui2011} introduce empirical null modeling into the approach of \cite{Efr09}
via an empirical Bayes method. They start with a similar $z$--score transform. However,
as a starting point absolute  values are used: 
\begin{gather}
z = \Phi^{-1}\left[1- 2\cdot\left\{1-F_{n_{l} + n_{k}-2}\left(\abs{\omega^{(k,l)}_{ \text{cat}}}
\right)\right\}\right] \,. \notag
\intertext{Additionally,  only a prior on the  absolute non-null 
effect sizes $g_A\left(\abs{\omega^{(k,l)}_{ \text{cat}}}\right)$ is assumed. 
The non--null $z$ have the conditional density}
f_A\left(z|\,\abs{\omega^{(k,l)}_{ \text{cat}}}\right)
= \varphi\left(\frac{\abs{\omega^{(k,l)}_{ \text{cat}}}-z}{V\left(\abs{\omega^{(k,l)}_{ \text{cat}}}\right)}\right) \, .\notag
\intertext{The variance function $V$ and the prior $g_A$ are 
estimated from the data.
As in \cite{Efr09} they also assume a two-group mixture model for the $z$--scores:}
f(z)  =  \eta_0 \varphi\left(\frac{z-\mu_0}{\sigma_0}\right) + (1-\eta_0) f_A(z) \, . \notag
\intertext{The null density is (in contrast to Efron) an empirical null, i.e. mean and
variance are estimated from the data: $f_0(z) = \varphi\left((z-\mu_0) /\sigma_0\right)$. 
The alternative density $f_A$ is computed as} 
f_A(z) = \int_0^{\infty} f_A\left(z|\abs{\omega^{(k,l)}_{ \text{cat}}}\right) 
g_A\left(\abs{\omega^{(k,l)}_{ \text{cat}}}\right)  d \abs{\omega^{(k,l)}_{ \text{cat}}} 
\notag\\
=\int_0^{\infty} \varphi\left(\frac{\abs{\omega^{(k,l)}_{ \text{cat}}}-z}
{\sqrt{V\left(\abs{\omega^{(k,l)}_{ \text{cat}}}\right)}}\right)
g_A\left(\,\abs{\omega^{(k,l)}_{ \text{cat}}}\right)  d\abs{\omega^{(k,l)}_{ \text{cat}}} \,.\notag
\intertext{The application of Bayes' rule gives a posterior expectation of
$\abs{\omega^{(k,l)}_{ \text{cat}}}$ which is unfortunately not as simple as  \eqcite{eq:conEff}:}
E_A\left(\abs{\omega^{(k,l)}_{ \text{cat}}}  | z\right) = 
\int_0^{\infty} \frac{f_A\left(z|\abs{\omega^{(k,l)}_{ \text{cat}}}\right) 
g_A\left(\abs{\omega^{(k,l)}_{ \text{cat}}}\right)}
{f_A(z)}  \, d \,\abs{\omega^{(k,l)}_{ \text{cat}}}  \notag \\
= \int_0^{\infty} \mspace{6.0mu}\abs{\omega^{(k,l)}_{ \text{cat}}} 
\frac{\varphi\Biggl(\frac{\abs{\omega^{(k,l)}_{ \text{cat}}}-z}{\sqrt{V\left(\abs{ 
\omega^{(k,l)}_{ \text{cat}}}\right)}}\Biggr)
g_A\left(\abs{\omega^{(k,l)}_{ \text{cat}}}\right)}{f_A(z)} \, d \abs{\omega^{(k,l)}_{ \text{cat}}} \, .\notag
\intertext{The statistic $\abs{\omega^{(k,l)}_{ \text{cat}}}$ is then
transformed back into a  absolute value effect size:}
E_A\left(\abs{\omega^{(k,l)}} | z\right)
= (1/n_{l}+1/n_{k})^{1/2}F_{n_{l} + n_{k}-2}^{-1}
\left(1- \frac{1}{2}\left[1-\Phi\left\{E_A\left(\abs{
\omega^{(k,l)}_{ \text{cat}}} | z\right)\right\}\right]\right) \, .\notag
%
%
\intertext{As in  \eqcite{eq:EffEf} the final effect size estimate is}
E\left(\abs{\omega^{(k,l)}} | z \right)
= E_A\left(\abs{\omega^{(k,l)}} | z\right)(1 - \fdr(z)) \, .\label{eq:Matfdr--effect} 
\end{gather}

In contrast to Efron's method the approach of \cite{Matsui2011} allows empirical 
null modeling and thus leads to better effect size estimates in general as \cite{Matsui2011}
also convincingly show in their article. 

However this increased accuracy comes at price. The estimation 
of variance function $V$ can take up to two hours. Furthermore it has to be estimated for
every number of class samples $n_k$ and $n_l$ separately. This makes cross-validation
based assessment of predictive accuracy extremely time
consuming. Additionally, even if $V$ has been computed for fixed  $n_k$ and $n_l$ 
the estimation of the final effect size will take up to a couple of minutes. 

In summary, while \cite{Matsui2011} provide a method that is superior 
to Efron's method in terms of bias, it also is computationally very demanding.

\subsubsection{A simple empirical-Bayes approach}\label{sec:simplEB}

In this section I will derive another more heuristic approach to the reliable estimation 
of effect sizes. It tries to combines the strengths of
\cite{Matsui2011} and \cite{Efr09}. 
Empirical null modeling will be included,  it will be computationally tractable and 
provide sufficient accuracy.    

Observe that in non-empirical Bayes frameworks reliable estimation of 
effect sizes
is  generally achieved by shrinking initial estimates of statistics playing the
same role as $\omega^{(k,l)}_{\text{cat}}$. 
For example in the popular PAM algorithm \citep{THNC03} the estimated $t$--scores
are shrunken using a parameter $\lambda$ estimated by cross validation. 

Therefore an appropriate adaptive shrinkage of the original should provide us
with reasonable effect size estimates. As it turns out, this adaptive shrinkage
can easily be achieved by employing false discovery rates.

The first step in my heuristic approach to achieve a shrinkage of $\omega^{(k,l)}$
is the assumption of a two component mixture model on the effect sizes:
\begin{gather}
f(\omega^{(k,l)}_{ \text{cat}})  =  \eta_0 f_0(\omega^{(k,l)}_{ \text{cat}}) 
+ (1-\eta_0) f_A(\omega^{(k,l)}_{ \text{cat}}) \,  ,\label{eq:groupE}
\end{gather}
leading to corresponding fdr estimates of \eqcite{eq:fdrDef}. 
Assuming a centered null distribution, we can now 
make use of the  ``naive'' estimates
$E_A\left(\omega^{(k,l)}\right) =  \omega^{(k,l)}$
 and correspondingly  
$E_0\left(\omega^{(k,l)} \right) =  0\, $ (since $f_0$ is centered).  
The 0 subscript indicates a conditioning on 
the null distribution, $E_0\left(\omega^{(k,l)} \right)
 = E\left(\omega^{(k,l)} \ | \omega^{(k,l)} \in \text
{``null''} \right)$. 
It now holds by the law of total probability and \eqcite{eq:fdrDef}
 that the effect  size is given by
\begin{align}
\begin{split}
E\left(\omega^{(k,l)} \right) 
&=  (1/n_{l}+1/n_{k})^{1/2} \biggl\{ E_0\left(\omega^{(k,l)}_\text{cat} \right) 
\cdot \prob\left(\omega^{(k,l)}_\text{cat} \in \text{``null''}  
| \omega^{(k,l)}_\text{cat} \right) \notag \\
&
\quad + 
E_A\left(\omega^{(k,l)}_\text{cat} \right) 
\cdot \prob\left(\omega^{(k,l)}_\text{cat} \in \text{``alternative''}  
| \omega^{(k,l)}_\text{cat} \right) \biggr\} 
\end{split}
\notag \\
&=(1/n_{l}+1/n_{k})^{1/2}E_A\left(\omega^{(k,l)}_\text{cat}  \right) 
\cdot \prob\left(\omega^{(k,l)}_\text{cat} \in \text{``alternative''}  
| \omega^{(k,l)}_\text{cat} \right) \notag \\
&=E_A\left(\omega^{(k,l)} \right) 
\cdot \left(1-\fdr( \omega^{(k,l)}_\text{cat} )\right) \notag \\
&= \omega^{(k,l)}\left(1-\fdr( \omega^{(k,l)}_\text{cat} )\right)
. \label{eq:fdr--effect} 
\end{align}
\eqcite{eq:fdr--effect} is very similar to \eqcite{eq:Matfdr--effect} and 
\eqcite{eq:unconEff}, however no full Bayesian posterior is 
computed. Instead simple non--Bayesian estimates for the expectations in the two groups 
model \eqcite{eq:groupE} are employed. This makes the implementation of 
\eqcite{eq:fdr--effect} computationally efficient.

There is an obvious downside though: 
Large (with respect to their absolute value) statistics usually have 
a high fdr value close to 1. Therefore they are hardly shrunken at all although 
their effect size is usually grossly overestimated. Thus it is necessary  to impose a minimum
shrinkage. From the results of the real data analysis in table 1 of  \cite{Matsui2011}
it can easily be seen that the empirical Bayes method  that these authors apply
imposes a shrinkage of at least 50\% on the top 5 test statistics. I therefore 
also set the minimum shrinkage
to 50\% leading to the formula
\begin{gather}
\bomega^{(k,l)}_{\fdr}
=  \bomega^{(k,l)} \cdot 
\min \left\{0.5; \, [1- \fdr( \omega^{(k,l)}_\text{cat} )]\right\} \, .
\label{eq:fdr--effect2} 
\end{gather}
I call this fdr--effect size estimation (fdr--effect) and 
abbreviate  $\omega^{(k,l)}\left(1-\fdr( \omega^{(k,l)}_\text{cat} )\right)$
 by  $\omega^{(k,l)}_{\fdr}$. Note that a fdr cutoff of 50\% is conceptually 
very close to Higher Criticism Thresholding \citep{KS2012}. 

Perhaps surprisingly, in next section it will 
be shown that it is competitive with regard to the attained accuracy even though
no sophisticated posterior estimates are used.  The adaptive shrinkage
performed in \eqcite{eq:fdr--effect2} can be interpreted as being in between 
the full empirical Bayes approaches of \cite{Efr09} or \cite{Matsui2011} 
and soft thresholding using a single shrinkage parameter for all
statistics as in \cite{THNC03}.

\subsection{Evaluation of Effect Size Estimation Methods on Real and Simulated Data}

\begin{figure}[t!]
\begin{center}
\centerline{\includegraphics[width=1\textwidth]{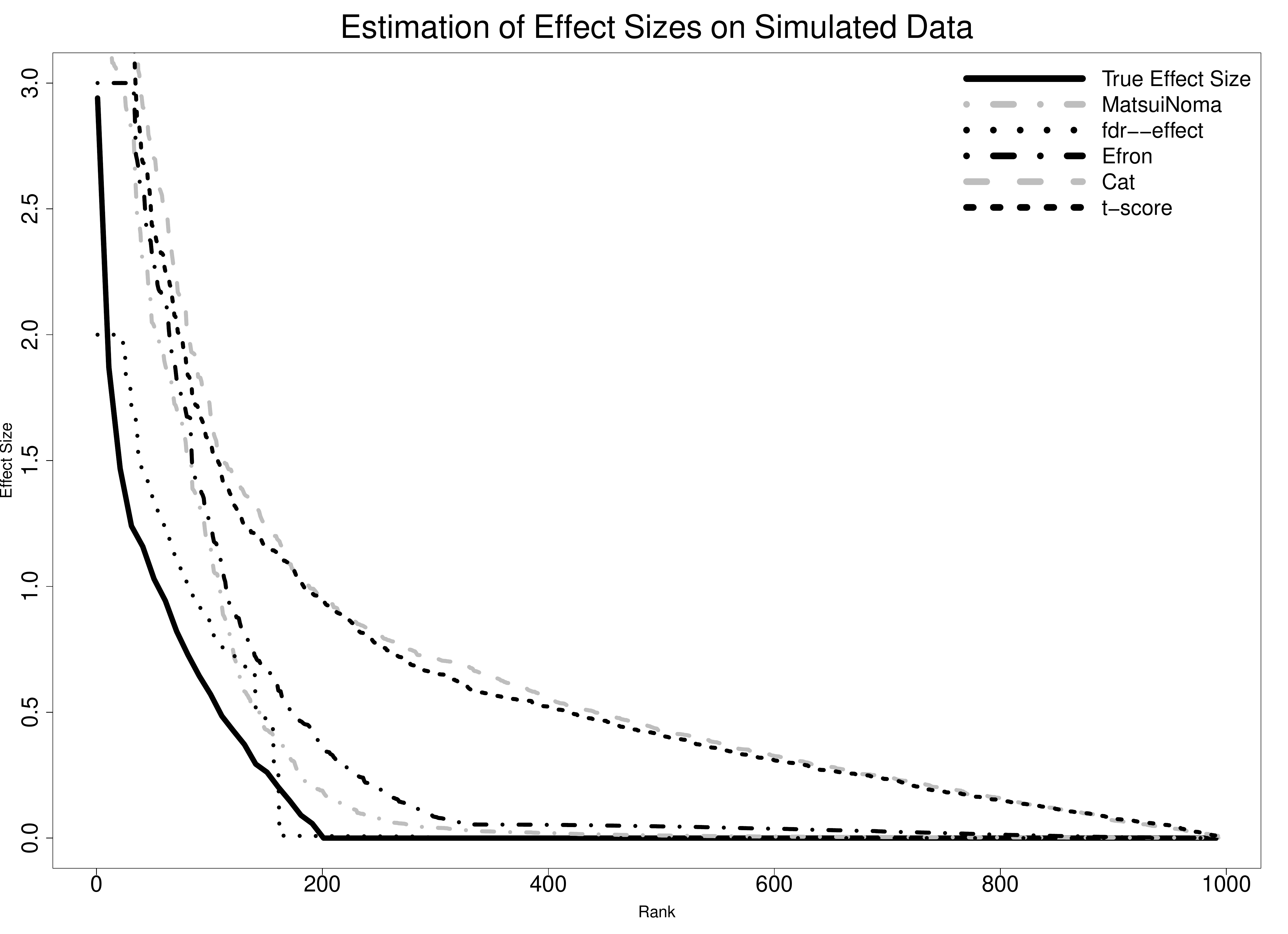}}
\caption{Comparison of  effect sizes on simulated data following the \citet{Smy04} model}
\label{fig:simulation}
\end{center}
\end{figure}

A comparison of effect size estimation methods using simulated data is shown in \figcite
{fig:simulation}. Specifically I compare the effect size 
estimation using  ``naives'' approaches  (simple cat and $t$  --scores) and the 
more sophisticated ones described in the previous section
 abbreviated as MatsuiNoma, Efron and 
fdr--effect respectively. For the methods  MatsuiNoma and Efron I use the implementations
offered by the authors, for fdr--effect  I perform  cat--score and fdr estimation 
using the \texttt{R}-packages \texttt{st} and  \texttt{fdrtool} \citep{Str08b}.
In the real data analysis displayed in \figcite{fig:singh} the package 
\texttt{locfdr} \citep{Efr04, Efr07b, Efr08a} 
is applied since this allows a straightforward use of an empirical
null as it has been suggested in  \cite{Matsui2011} 
and \cite{Efr04} for this data set.

I follow closely the setup used in \cite{Smy04}, \cite{OS07a} and \cite{ZS09}
to simulate gene expression data. 
The parameters are chosen in such a way that  effect sizes between 1 and 3 are obtained
which roughly corresponds to the range considered in the simulation studies of \cite{Matsui2011}.

The number of statistics was fixed at $d=1000$ with 200 statistics designated to be 
differentially expressed. The variances across genes were drawn from a  scale--inverse--chi--
square distribution  $\text{Scale-inv-}\chi^2(d_0, s_0^2)$ with $s^2_0=1$ and $d_0=1$, 
i.e. the variances vary moderately from gene to gene. Furthermore, the difference of means 
for the  differentially expressed genes (1--200) were drawn from a  normal distribution 
with mean  zero and the gene-specific variance multiplied with a scale factor  set to 
$0.3$.  For the non--differentially expressed genes (201--1000) the difference was set to 
zero. The data were generated by drawing from group-specific multivariate normal 
distributions with the given variances and means, employing a block diagonal correlation 
structure intended to mimic gene expression data. This structure was  generated as 
in \citet{GHT07} with block size  100 and block entries equal to $0.9^{\abs{i-j}}$. 
Furthermore, the sample sizes $n_1$ and $n_2$ are equal with $n_1=n_2=8$.

The effect size estimates  are plotted in \figcite{fig:simulation}
according to their rank. It is important to note that this does not tell us whether the 
respective ranking is correct. Thus, even though  the effect size estimates of 
the cat--score and an ordinary $t$--score are very similar, this does not mean that 
their induced ranking is comparable. 

It can be seen that fdr--effect and MastsuiNoma yield good results, while Efron's method 
has a higher bias for effect sizes up to 1, a phenomenon already observed by \cite
{Matsui2011}.
The ``naive'' approach using cat--scores is far off for effect sizes up to 1.5. 
However all methods overestimate large effect sizes. It  follows  that variable 
selection methods relying on effect size estimates will generally have a tendency
of choosing only a relatively  small number of variables in data sets with 
large effects.

\begin{figure}[t!]
\begin{center}
\centerline{\includegraphics[width=1\textwidth]{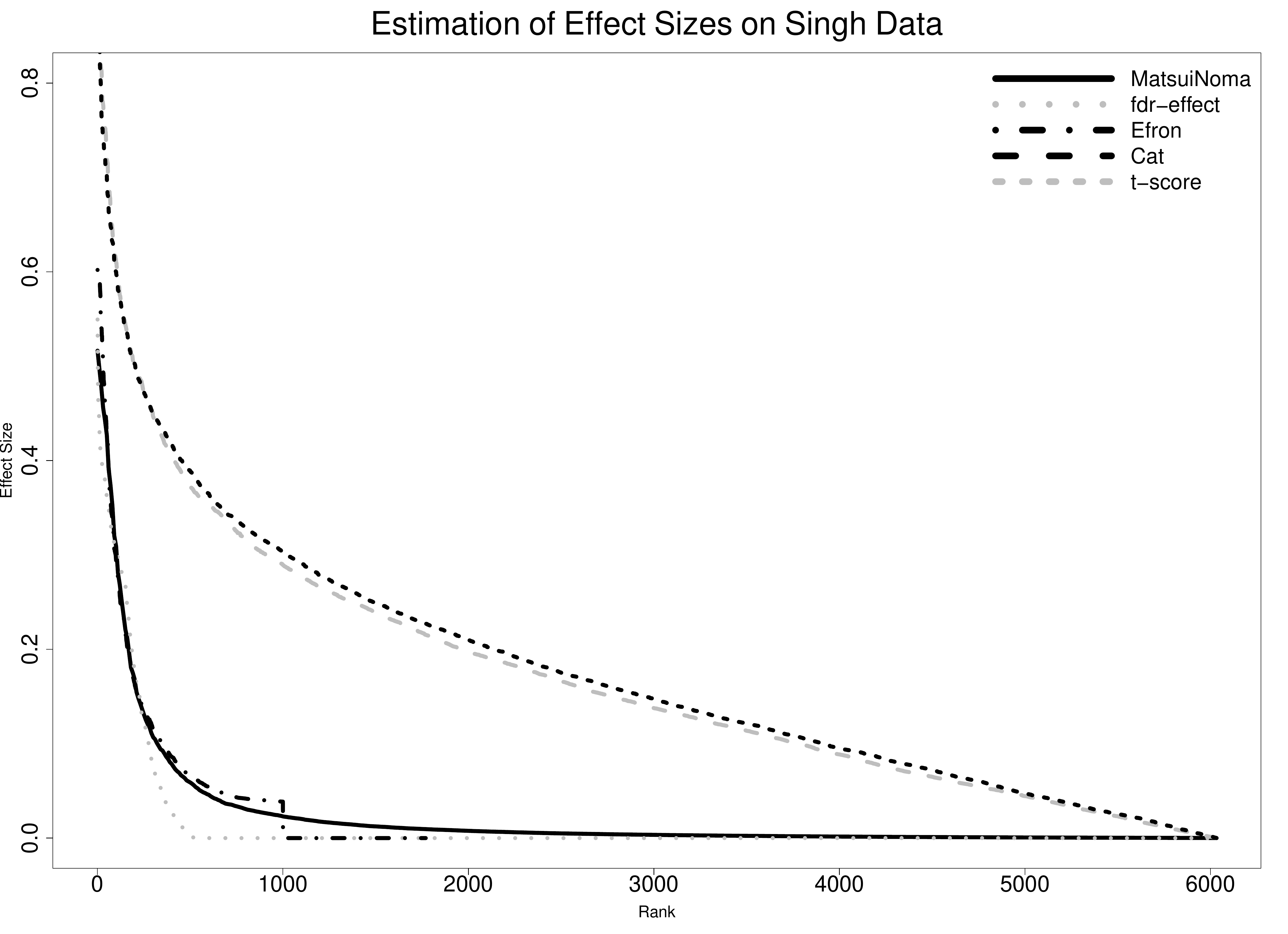}}
\caption{Comparison of  effect sizes for the \cite{SF+02} data}
\label{fig:singh}
\end{center}
\end{figure}

This is in fact a phenomenon already observed by  \cite{AS2010} for the Efron algorithm
applied to the \cite{SF+02} prostate cancer gene expression data. 
This data consists of gene expression measurements
of $d=6033$ genes for $n=102$ patients, of which 52 are cancer
patients and 50 are healthy. It has already been analyzed in 
\cite{Efr09} and \cite{Matsui2011}. \figcite{fig:singh} shows
the analysis results. As in the simulated data the ``naive'' approaches
are far off, while Efron and MatsuiNoma are quite similar. Note however, that
MatsuiNoma gives significantly lower estimates of large effect sizes than Efron
a phenomenon already noted in \cite{Matsui2011}. The fdr--effect method
yields similar results to MatsuiNoma for large effect sizes but arrives 
at zero estimates much faster than MatsuiNoma and Efron. In conclusion 
all empirical Bayes methods considered seem to give sound results here,
while the empirical methods are probably grossly overestimating the effect
sizes.

\section{Variable Selection and Estimation of the Prediction Rule}
\label{VAR}
\subsection{Estimation of the prediction rule and local false discovery rates}

For the estimation of the prediction rule  (\eqcite{eq:multiclasslda}) I mostly employ 
James-Stein-type estimators as in shrinkage discriminant analysis  -- SDA, \cite{AS2010}. 
The group centroids $\bmu_k$ are estimated by the empirical means, for the correlations $\bP$ 
the ridge-type estimator from  \citet{SS05c} is used and the  variances $\bV$ are estimated by 
the shrinkage estimator from \citet{OS07a}. Finally  the proportions $\pi_k$ are obtained by 
using the frequency estimator from \citet{HS09}. For SDA I employ the implementation provided by
the \texttt{R} package \texttt{sda}. The local false discovery rates used in
the fdr--effect approach are learned by  using the Grenander density
estimator and truncated maximum likelihood for the empirical null as in \cite{Str08c}.
As in chapter  \ref{effect} the implementation offered by the \texttt{R} 
package \texttt{fdrtool} is employed.

\subsection{Variable ranking and selection}

\subsubsection{Variable ranking}
Before being able to select variables a variable ranking needs to be established (obj. 
(i)). In the two class case this is straightforward since the feature weight vector
for class one $\bomega_1$ corresponds to the effect size vector $\bomega^{(1,2)}$ and the 
feature weight vector for class two $\bomega_2$ to the effect size vector
$-\bomega^{(1,2)}$. Thus variables can be ranked according to the absolute value of $
\bomega^{(1,2)}$. In the the case of multiple classes the situation is more complicated.
The feature weight vectors of the different classes need to be summarized in a certain
way to obtain the importance of each feature $i$ in class prediction. Here 
I use the summary statistic $S_i$ proposed by \citep{AS2010} and given by

\begin{gather}
S_i = \sum_{k=1}^K \left(\omega_{\text{cat}, i}^{(k, \text{pool})}\right)^2 \; ,
\label{eq:summaryscore}
\end{gather}
where 
$\omega_{\text{cat}, i}^{(k, \text{pool})} 
= (1/n_{k}-1/n)^{-1/2}  \omega_{i}^{(k, \text{pool})} $.  
Since false discovery rates are generally assumed to be monotone, 
\eqcite{eq:fdr--effect} shows that 
using fdr--effect effect size estimates $\bomega^{(k,\text{pool})}_{\fdr}$  would produce
the same ranking as the  cat--scores if they
were used instead of $\bomega^{(k,\text{pool})}_{\text{cat}}$ 
to compute $S_i$ in \eqcite{eq:summaryscore}.

\subsubsection{Misclassification rate based variable selection}
Having obtained estimates  $\widehat \bomega^{(k,l)}_{\fdr}$ of $\bomega^{(k,l)}_
{\fdr}$ and $\widehat \pi_{k}$ of $\pi_{k}$ we can now compute an estimate of the
missclassification rate using \eqcite{eq:terror}.
Let $\widehat \bomega^{(k,l)}_{\fdr}(t)$ be the vector of the $t$ top-ranked variables
according to the ranking induced  by the vector $\bS$ of all statistics 
$S_i$ given by \eqcite{eq:summaryscore}.
We then have an estimate of the misclassification rate which depends on $t$:
\begin{gather}
\widehat{ \prob }(\text{error})(t)
= \sum_{k=1}^{K} \widehat \prob(  \neq k | k)(t)  \times \widehat \prob(  k) \notag \\
= \sum_{k=1}^{K}  \Phi \biggl( - \min_{l \neq k} 
\frac{[\widehat \bomega^{(k,l)}_{\fdr}(t)]^T [\widehat 
\bomega^{(k,l)}_{\fdr}(t)] + 
2 \log\biggl(\frac{\pi_k}{\pi_l}\biggr) } 
{2 \sqrt{[\widehat \bomega^{(k,l)}_{\fdr}(t)]^{T}[\widehat 
\bomega^{(k,l)}_{\fdr}(t)]}}  \biggr) \times \widehat \pi_{k} \, . \label{eq:terrorVS}
\end{gather}

\cite{Efr09} performs feature selection by choosing a level $\alpha = 0.05$ 
as a target missclassification rate for the estimate in \eqcite{eq:terrorVS}. 
Although one could view $\alpha$ as a tuning parameter I follow his suggestion
in this regard since experiments with lower results only lead to very large
feature sets showing only to a negligible improvement of the classification performance.  

After the target error $\alpha$ has been set a feature threshold $t^*$ is 
obtained by including as many features as necessary to reach it, i.e.
$\widehat{ \prob }(\text{error})(t^*) = \alpha$. 
Since usually a lot of features are shrunken to zero, it is possible that
the target error can not be reached. Then all the features will be included.
This however is extremely unlikely to happen in real high dimensional data analysis.
Finally all features fulfilling $S_i \geq S_t^*$ 
are included in the classifier.

\section{Analysis of Real and Simulated Data }
\label{RES}
\subsection{Simulations}
In this section I compare variable selection based on the misclassification rate
(MR)
 with several other state of the art thresholding variable selection 
approaches, namely false-non discovery
rate (FNDR) thresholding  \citep{AS2010}, 
Higher Criticism (HC) thresholding  \citep{DJ08} 
and the PAM algorithm  \citep{THNC03}. As a base line classifier I also
include the results of classification with all features, i.e. performing
no variable selection. 

The simulations follow closely the setup of \cite{WT11}. A training
set of size 100 and  a test set of 1000 samples are created with a dimension of 
$d =500$ variables. In total 25 runs of each simulation setup are performed.

\subsubsection{Simulation setup 1}
In this setup there are four classes with equal probability (0.25) no
correlation and unit variance.  25 features are differentially expressed 
in each class with an effect size of 0.7, yielding a total number of 100 differentially 
expressed features. Since there is no correlation we perform Diagonal Discriminant
Analysis (DDA), i.e. LDA with identity covariance $\bSigma = I_d$. The results are 
displayed in
\tabcite{Sim1}.

\begin{table}[!t]

\caption{Prediction errors and number of selected features for  simulation setup 1,
the number in the round brackets is the estimated standard error over 25 runs.
the true number of differentially expressed features is 100.}
\begin{center}
\begin{tabular}{lrr}
\toprule
Method       &  Prediction Error &  Features \\
\midrule
DDA-MR         & 0.1077 (0.0177) & 156.48 (\phantom{0}64.70) \\
DDA-FNDR       & 0.2482 (0.1272)  & \phantom{0}39.24 (\phantom{0}23.72)\\
DDA-HC         & 0.1880 (0.0626)  & 152.32 (193.48) \\
PAM            & 0.0923 (0.0163) & 253.6\phantom{0} (116.26)  \\
DDA-ALL        & 0.1555 (0.0180) & 500\phantom{.0 } \phantom{(000.00)}\\
\bottomrule
\end{tabular}
\label{Sim1}
\end{center}
\end{table}

It can be seen that thresholding the
summary statistic $S$ (\eqcite{eq:summaryscore}) by false-non discovery
rates or Higher Criticism  yields hardly any  significant features in most runs. 
Consequently the estimated prediction errors are quite high.

Misclassification rate based feature selection as well as PAM 
however identify  features useful for classification. This indicates
that ''analytical`` thresholding methods, which do not rely on the optimization
of a tuning parameter may not work reliably when the effect
sizes are small.

\subsubsection{Simulation setup 2}
In this simulation I use a \cite{GHT07} type block correlation with 5 blocks of
size $100 \times 100$.  As in section \ref{effect} each  
block entry is given by  $0.9^{\abs{i-j}}$, thus we have some highly 
correlated variables within blocks but variables in different blocks are independent. 

Note that  \cite{WT11} report to use an entry size of
0.6. This  is probably a  misprint
since my results obtained for PAM are quite similar to the 
ones reported in their article, 
while for 0.6 the error of PAM only about 5 \%.

There are two classes with equal probability (0.5) and 200 features are 
differentially expressed with effect size 0.6, all of them
are attributed to class 2. Since there is correlation in this setting 
I perform LDA.

\begin{table}[t]

\caption{Prediction errors and number of selected features for  simulation setup 2,
the number in the round brackets is the estimated standard error over 25 runs. The true
number of differentially expressed features is 200.}
\begin{center}
\begin{tabular}{lrr}
\toprule
Method       &  Prediction Error &  Features \\
\midrule
LDA-MR     	& 0.000 (0.000) & 63.16 (7.215) \\
LDA-FNDR    	& 0.000 (0.000) & 60.96 (6.567) \\
LDA-HC  	& 0.000 (0.000) & 85.04 (8.677) \\
PAM             & 0.088 (0.018) & 294.0 (69.43)\\
LDA-ALL         & 0.093 (0.014) & 500\phantom{.0} \phantom{(00.00)}\\
\bottomrule
\end{tabular}\\
\label{Sim2}
\end{center}
\end{table}

It can be seen  in \tabcite{Sim2} that all feature selection methods except for 
PAM, which does not take  correlation into account, perform quite well here.

\subsection{Gene expression data}

In \citet{AS2010} the relative effectiveness of 
the FNDR and HC thresholds
to select relevant genes in shrinkage discriminant analysis
applied to gene expression data has already been compared. 
I followed their setup here and analyzed four clinical
gene expression data
sets related to prostate cancer \citep{SF+02}, B-cell
lymphoma \citep{AED+00}, colon cancer \citep{AB+99}, 
and brain cancer \citep{PTG+02}. 

Specifically, 
balanced 10-fold cross-validation with 20 repetitions was performed to
obtain error estimates and their standard deviations.
The number of selected features is inferred by single run of the
respective variable selection method on the whole data set. Only for PAM
this was repeated several times, since the number of selected
variables selected by this algorithm varies considerably between
several runs in a row on the same data set.

In \tabcite{tab:cancersda}
it can bee seen that my approach has a performance similar to the other approaches.
Interestingly, the MR approach shows a more ``adaptive'' feature selection, leading
to appropriate feature sets for each problem. In the brain data set 
a very compact set of features is selected yielding a prediction
error which is nonetheless in the range of the other approaches. The same
is true for the Lymphoma and Colon data sets. This demonstrates that
a variable selection method based on effect sizes leads to compact and yet
effective molecular signatures.   

\begin{table}[!t]
\caption{Analysis of four cancer gene expression data sets with shrinkage 
discriminant analysis. The number of selected features are determined by
a single feature selection run on the whole data set.}
\begin{center}
\begin{tabular}{lrr}
\toprule
Data / Method  &  Prediction Error & Selected Variables \\
\midrule
\midrule
\multicolumn{3}{l}{ {\bf Prostate} ($d=6033,n=102, K=2$) } \\
LDA-MR         & 0.0630 (0.0050) & 134 \\
LDA-FNDR       & 0.0550 (0.0048) & 131 \\
LDA-HC         & 0.0497 (0.0045) & 116 \\
PAM  	       & 0.0850 (0.0061) & 172-377 \\
\midrule
\multicolumn{3}{l}{ {\bf Lymphoma } ($d=4026,n=62, K=3$) }\\
LDA-MR       & 0.0211 (0.0039) & 34 \\
LDA-FNDR     & 0.0036 (0.0018) & 392 \\
LDA-HC       & 0.0000 (0.0000) & 345 \\
PAM 	     & 0.0234 (0.0041) &  2796 -- 2383\\

\midrule
\multicolumn{3}{l}{ {\bf Colon} ($d = 2000, n=62, K=2$) } \\
LDA-MR      & 0.1291  (0.0093) & 28 \\
LDA-FNDR    & 0.1278  (0.0088)  & 168 \\
LDA-HC      & 0.1233  (0.0087) & 122 \\
PAM 	    & 0.1160  (0.0921) & 13-23\\
\midrule
\multicolumn{3}{l}{ {\bf Brain} ($d=5597,n=42, K=5$) } \\
LDA-MR       & 0.1628   (0.0126) &  56  \\
LDA-HC       & 0.1417   (0.0108) & 131 \\
LDA-FNDR     & 0.1525   (0.0120) & 102 \\
PAM   	     & 0.2023   (0.0118) & 42--5587\\

\bottomrule 
\end{tabular}\\
\end{center}
\label{tab:cancersda}
\end{table}

\section{Discussion}

In this paper I reviewed and extended statistical techniques related to effect 
size estimation in linear classification and showed how to use them for variable
selection. The fdr--effect method proposed for effect size estimation  has
been shown to work as 
well as competing approaches while being conceptually
simple and computationally inexpensive. It therefore successfully unites the strengths
of the approaches presented in  \cite{Efr09} and \cite{Matsui2011}.

Additionally, I gave a unified treatment
of the effect size estimation approaches presented in these two papers
elucidating similarities not apparent when considering the original publications 
only.

 Variable selection  by minimizing 
the misclassification rate has been somewhat neglected in the literature but
I showed in accordance with \cite{DS07}, \cite{Efr09} and \cite{Matsui2011} that 
it is indeed very well suited for real world problems.  In addition, 
it is also much more intuitive than selecting a non-interpretable
regularization parameter as for example in the PAM algorithm and leads
to compact and interpretable feature sets.

In this work I proposed a conceptually simple and competitive variable
selection  algorithm that gives priority to genes with large effect sizes and
is thus easy to interpret. This has been achieved 
by extending and combining the ideas of \cite{DS07}, \cite{Efr09} and
\cite{Matsui2011}.  

High expectations are associated with the promise of a personalized medicine 
promising tailored treatments based on genetic and other information of the patient.
In order to develop molecular diagnostics guiding  these treatments,
statistical approaches for effective and interpretable classification are indispensable.

The methodology presented in this article provides interpretability 
and applicability for biological study and medical use. Reliable 
effect size estimates allow one to identify genes having discriminative 
power while variable selection based on these effect size estimates
allows the selection of the most important genes for the construction of
classification algorithms.

%% file: acknowledgments.tex
\section*{Acknowledgments}
I would like to thank  Shigeyuki Matsui for providing \texttt{R}--Code 
implementing the method of \cite{Matsui2011}. Furthermore, I thank 
St\'ephane Robin, Tristan Mary-Huard, Marie-Laure Martin-Magniette (all at
AgroParisTech) and Korbinian Strimmer, David Petroff 
as well as Verena Zuber (all at University of Leipzig)  
for fruitful discussions of this work.  Korbinian Strimmer  and  Verena Zuber also
provided \texttt{R}-Code implementing the simulation setup of \citet{Smy04}.
I also thank Miika Ahdesm\"aki (Almac Diagnostics) for \texttt{R}-Code performing
CV--based prediction error estimation of several classification methods.